\documentclass{PoS}
\usepackage{amsmath}
\usepackage{amssymb}
\usepackage{graphicx}

\title{Exploring the phase diagram of finite density QCD 
at low temperature by the complex Langevin method}

\ShortTitle{Exploring the phase diagram of finite density QCD 
at low temperature by the CLM}

\author{\speaker{Yuta Ito}\thanks{KEK-TH/2087}\\
        High Energy Accelerator Research Organization (KEK),\\
        1-1 Oho, Tsukuba, Ibaraki 305-0801, Japan\\
        E-mail: \email{yito@post.kek.jp}}

\author{Hideo Matsufuru\\
        High Energy Accelerator Research Organization (KEK),\\
        1-1 Oho, Tsukuba, Ibaraki 305-0801, Japan\\
	E-mail: \email{hideo.matsufuru@kek.jp}}

\author{Jun Nishimura\\
        High Energy Accelerator Research Organization (KEK),\\
	1-1 Oho, Tsukuba, Ibaraki 305-0801, Japan, and\\
	Graduate University for Advanced Studies (SOKENDAI),\\
        1-1 Oho, Tsukuba 305-0801, Japan\\
	E-mail: \email{jnishi@post.kek.jp}}

\author{Shinji Shimasaki\\
	Research and Education Center for Natural Sciences,
	Keio University, Hiyoshi 4-1-1, Yokohama, Kanagawa 223-85
	21, Japan
	E-mail: \email{shinji.shimasaki@keio.jp}}

\author{Asato Tsuchiya\\
	Department of Physics, Shizuoka University,
	836 Ohya, Suruga-ku, Shizuoka 422-8529, Japan \\
	E-mail: \email{tsuchiya.asato@shizuoka.ac.jp}}

\author{Shoichiro Tsutsui\\
        High Energy Accelerator Research Organization (KEK),\\
        1-1 Oho, Tsukuba, Ibaraki 305-0801, Japan\\
        E-mail: \email{stsutsui@post.kek.jp}}

\abstract{Monte Carlo studies of QCD at finite density 
suffer from the sign problem, 
which becomes easily uncontrollable as the chemical potential $\mu$ 
is increased
even for a moderate lattice size.
In this work we make an attempt to approach the high density 
low temperature region by the complex Langevin method (CLM)
using four-flavor staggered fermions with reasonably small quark mass
on a $8^3 \times 16$ lattice. 
Unlike the previous work on a $4^3 \times 8$ lattice, 
the criterion for correct convergence is satisfied 
within a wide range of $\mu$ without using the deformation technique. 
In particular, the baryon number density exhibits a plateau behavior
consistent with the formation of eight baryons,
and it starts to grow gradually at some $\mu$.}

\FullConference{The 36th Annual International Symposium 
on Lattice Field Theory - LATTICE2018\\
		22-28 July, 2018\\
		Michigan State University, East Lansing, Michigan, USA.}

\begin{document}

\section{Introduction}

In finite density QCD at low temperature and high density,
it is conjectured that there are various interesting phases 
such as the quark matter phase, the color-superconductor phase and so on.
Exploring these phases based on full QCD simulation is important 
not only for purely academic reasons but also for
understanding the interior structure of neutron stars.
However, standard Monte Carlo methods face with
the sign problem, 
which has been hindering the development in this direction so far.

The complex Langevin method (CLM) \cite{Klauder:1983sp,Parisi:1984cs}
has been studied as a promising approach to this problem.
The idea is to extend the stochastic quantization
based on the Langevin equation to the cases with a complex action
by complexifying the dynamical variables,
where holomorphicity has to be respected in 
defining the drift term and the observables.
The biggest problem of this approach is that
the obtained results are not guaranteed to be correct.
However, in recent years the reasons for 
the failure have been 
clarified \cite{Aarts:2009uq,Nishimura:2015pba,Nagata:2016vkn}, 
and various techniques \cite{Aarts:2009dg,Seiler:2012wz,Ito:2016efb}
have been developed
to extend the applicability of this method.
Thanks to these developments, 
finite density QCD in the heavy dense 
limit \cite{Seiler:2012wz,Aarts:2013uxa,Aarts:2016qrv}
and at high temperature \cite{Sexty:2013ica,Fodor:2015doa}
has been studied successfully.

We attempt to extend this success to
the low temperature region with reasonably small quark mass,
where the transition to the nuclear matter phase 
and subsequently to the quark matter phase
at larger chemical potential $\mu$ is anticipated.
Extending the previous work \cite{Nagata:2018mkb}
on a $4^{3}\times 8$ lattice,
we perform simulations with 
four-flavor
staggered fermions 
on a $8^{3}\times 16$ lattice with quark mass $m=0.01$ and $m=0.05$.
(See refs.~\cite{Sinclair:2017zhn}
for related work with two-flavor staggered fermions.)
Rather surprisingly, we find that
the criterion for correct convergence \cite{Nagata:2016vkn}
is satisfied within a wide range of $\mu$ without using the
deformation technique \cite{Ito:2016efb}
unlike the previous work \cite{Nagata:2018mkb}.
In particular, the baryon number density exhibits a plateau
as a function of $\mu$, which is
consistent with the formation of eight baryons,
and it starts to grow gradually at some $\mu$.
We obtain similar results for the two values of the quark mass
although the plateau behavior becomes clearer for $m=0.01$.

We also simulate the phase quenched model 
with the same set of parameters using the standard RHMC algorithm.
Here we obtain quite different results
for the two values of the quark mass $m$ unlike the results for the full model.
At $m=0.01$, the baryon number density exhibits a different plateau
with lower height at smaller values of $\mu$, which is consistent with
the formation of four mesons considering that 
$\mu$ in the phase quenched model actually corresponds to
the isospin chemical potential.
Thus we find a clear difference between the results of the CLM and
those for the phase quenched model at $m=0.01$.
At $m=0.05$, on the other hand, 
the baryon number density behaves quite similarly
for the two models.

The rest of this paper is organized as follows. In section \ref{sec:2}
we explain briefly how we apply the CLM to finite density
QCD and how we judge the validity of the results. 
In section \ref{sec:Results}
we show the results obtained by the CLM and compare them
with the results for the phase quenched model.
The section \ref{sec:Summary-and-discussions}
is devoted to a summary and discussions. 

\section{Complex Langevin method for finite density QCD
\label{sec:2}
}

In this paper we study finite density QCD with
$N_{\mathrm{f}}=4$ staggered fermions. The partition function is
given as
\begin{equation}
Z=\int dU\,\det M\left[U,\mu\right]e^{-S_{\rm g}[U]} \ ,
\end{equation}
where $U_{x,\nu}\in {\rm SU}(3)$ are the link variables
and $S_{\rm g}[U]$ is the plaquette action defined by
\begin{equation}
S_{\rm g}[U]=-\frac{\beta}{6}\sum_{x}\sum_{\nu \neq \rho}
         \mathrm{tr}(U_{x,\nu}U_{x+\hat{\nu},\rho}
U_{x+\hat{\rho},\nu}^{-1}U_{x,\rho}^{-1}) \  .
\end{equation}
The determinant $\det M[U,\mu]$, which is obtained by integrating
out the fermions, is complex when the chemical potential $\mu$ is nonzero. 
%
This makes the standard Monte Carlo simulation difficult
due to the sign problem, which becomes severer as $\mu$ is increased. 

In order to overcome this problem, we apply the CLM,
which is a complex extension of the stochastic quantization 
based on the Langevin equation. In this method, 
the link variables $U_{x,\nu}$ are
complexified as $\mathcal{U}_{x,\nu}\in {\rm SL}(3,C)$, and accordingly
the drift term and the observables, which are functions of $U_{x,\nu}$,
have to be extended to functions of $\mathcal{U}_{x,\nu}$ 
holomorphically.
The complexified link variables are updated according to the complex version
of the Langevin equation
\begin{equation}
\mathcal{U}_{x,\nu}(t+\epsilon)=
\exp\left[i\left(-\epsilon v_{x,\nu}(\mathcal{U}(t))
+\sqrt{\epsilon}\eta_{x,\nu}(t)\right)\right]
\mathcal{U}_{x,\nu}(t) \ ,
\label{eq:cle}
\end{equation}
where $t$ is the discretized Langevin time
and $\epsilon$ is the stepsize. 
The drift term $v_{x,\nu}(\mathcal{U})$ in eq.~(\ref{eq:cle})
is defined by the holomorphic extension of
\begin{align}
 v_{x,\nu}(U)
= \sum_a  \lambda_a
\left.\frac{d}{d\alpha}
S(e^{i \alpha \lambda_a}U_{x ,\nu})\right|_{\alpha=0}  \ ,
\label{drift}
\end{align}
where 
$S[U]=S_{\rm g}[U]-\ln\det M[U,\mu]$
and 
$\lambda_a \ (a=1,\cdots,8)$ are
the SU(3) generators normalized 
by $\mathrm{tr}(\lambda_a \lambda_b) = \delta_{ab}$.
The noise term $\eta_{x,\nu}(t)$ in eq.~(\ref{eq:cle}), which
are $3\times3$ traceless Hermitian matrices, are generated with the
Gaussian distribution 
$\exp (-\frac{1}{4}\mathrm{tr}\{\eta_{x,\nu}^{2}(t)\})$.

The expectation value of an observable $O(U)$
can be obtained as
\begin{equation}
\langle O(\mathcal{U})\rangle=
\lim_{T\rightarrow\infty}\frac{1}{T}\int_{t_{0}}^{t_{0}+T}dt\,
\langle O(\mathcal{U}(t)) \rangle_{\eta} \ ,
\end{equation}
where the expectation value 
$\langle \ \cdot  \  \rangle_{\eta}$ 
appearing on the right-hand side
should be taken with respect to the
Gaussian noise $\eta$, 
and $t_{0}$ should be sufficiently large to achieve thermalization. 
The effect of the complex fermion determinant is
included in the complex Langevin process (\ref{eq:cle}) through the
drift term, and there is no need for reweighting unlike 
the path-deformation approach such as 
the generalized Lefschetz-thimble method.

It is known that the CLM fails to yield correct results in some cases. 
However, we can judge whether the obtained results are
correct or not in the following way \cite{Nagata:2016vkn}. 
Let us define the magnitude of the drift term as
\begin{equation}
v= \sqrt{\frac{1}{3} 
\max_{x, \nu}\mathrm{tr}(v_{x,\nu}^{\dagger} v_{x,\nu})} \ ,
\label{eq:drift_norm}
\end{equation}
and consider its probability distribution $p(v)$.
If $p(v)$ falls off exponentially or faster, the result is reliable. 
A slower fall-off such as a power law
can be caused either by the excursion problem \cite{Aarts:2009uq}
or the singular-drift problem \cite{Nishimura:2015pba}.
The first problem occurs
when the link variables $\mathcal{U}_{x,\nu}$ become too far
from SU(3) matrices, and the second one occurs when the Dirac
operator has near-zero eigenvalues frequently
because the drift term involves 
the inverse of the Dirac operator.

In order to suppress the excursion problem as much as possible,
we perform the gauge cooling \cite{Seiler:2012wz} as follows.
(See ref.~\cite{Nagata:2016vkn} for justification.)
Let us define the unitarity norm as
\begin{equation}
{\cal N}=\frac{1}{12 N_{\rm V}}\sum_{x,\nu}
\mathrm{tr}(\mathcal{U}_{x,\nu}^{\dagger} \mathcal{U}_{x,\nu}-{\bf 1})\ ,
\end{equation}
where $N_{\rm V}$ is the number of lattice sites.
This quantity 
measures how far $\mathcal{U}_{x,\nu}$ is from SU(3) configurations,
and it vanishes if and only if
all of $\mathcal{U}_{x,\nu}$ are unitary matrices.
After updating $\mathcal{U}_{x,\nu}$ by the complex Langevin equation
(\ref{eq:cle}), we perform a complexified gauge transformation
\begin{equation}
\mathcal{U}_{x,\nu}\rightarrow 
g_{x}\mathcal{U}_{x,\nu}g_{x+\hat{\nu}}^{-1} \ , \quad
\mathrm{where~} g_{x}\in {\rm SL}(3,C) 
\end{equation}
in such a way that the unitarity norm is minimized.


\begin{figure}[t]
\begin{centering}
\includegraphics{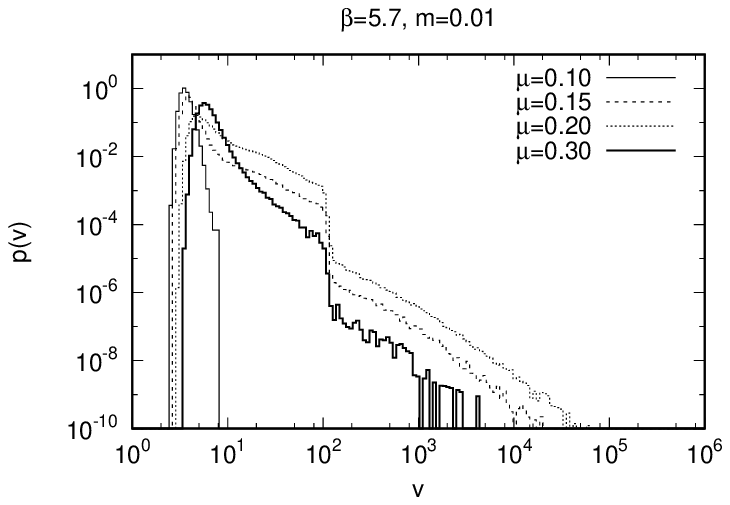}\includegraphics{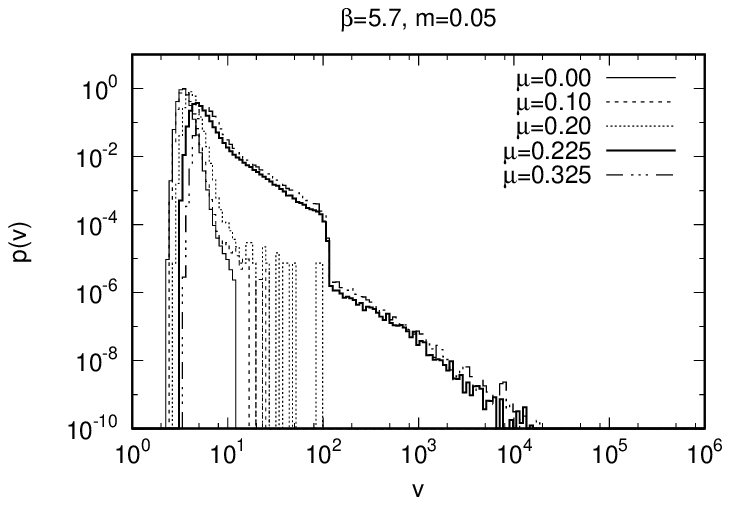}
\par\end{centering}
\begin{centering}
\includegraphics{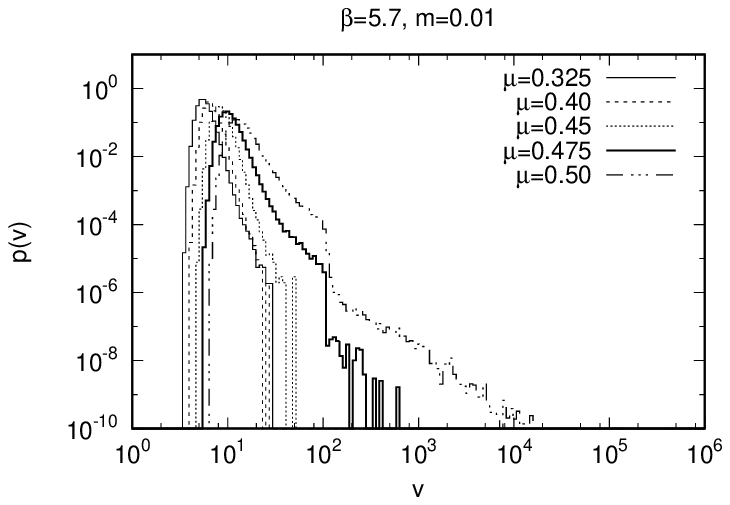}\includegraphics{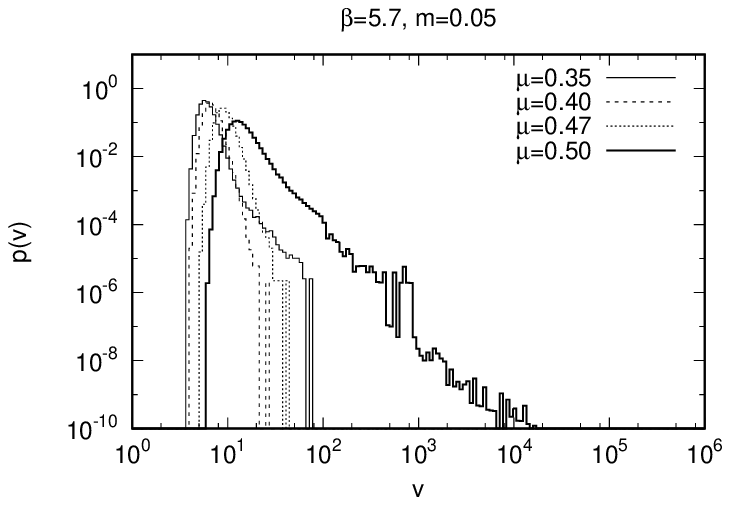}
\par\end{centering}
\caption{The probability distribution $p(v)$ of the drift term are plotted
for various $\mu$ with $m=0.01$ (Left) and $m=0.05$ (Right). 
\label{fig:drift_histogram}}
\end{figure}

\section{Results\label{sec:Results}}

We have performed simulations on a $8^{3}\times16$ lattice with $\beta=5.7$
and the quark mass $m=0.01$ and $0.05$. 
The Langevin stepsize is chosen initially as $\epsilon = 10^{-4}$, which
is reduced adaptively when the magnitude of the drift exceeds 
certain threshold \cite{Aarts:2009dg}.
We have made $(5 \sim 15 ) \times 10^{5}$ total Langevin steps 
for each set of parameters.

Let us first check the validity of the results obtained by the CLM.
In Fig.\ \ref{fig:drift_histogram} (Left), the probability distribution
$p(v)$ of the drift term is plotted for various $\mu$
with $m=0.01$. 
From these plots, we find that $p(v)$ 
for $\mu = 0.1$ and
$0.325 \leq \mu \leq 0.475$ 
shows a clear exponential
fall-off and therefore the simulations are reliable. On the other
hand, $p(v)$ for $0.15 \leq\mu\leq0.3$ and $\mu=0.5$ shows 
a power law fall-off, which is actually caused by
the contribution from the fermion determinant. 
Thus, we find that the singular-drift problem occurs 
in these parameter regions. 
Similarly, in Fig.\ \ref{fig:drift_histogram} (Right), we show the
probability distribution $p(v)$ 
for various $\mu$ with $m=0.05$,
from which we find that the results for
$0\leq\mu\leq0.2$ and $0.35\leq\mu\leq0.47$ 
are reliable. As in
the $m=0.01$ case, the power law fall-off seen in these plots
is due to the singular-drift problem.

\begin{figure}
\begin{centering}
\includegraphics{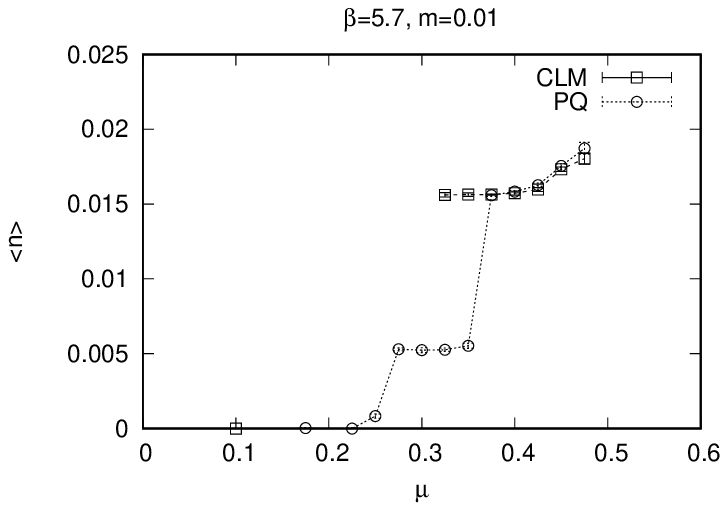}\includegraphics{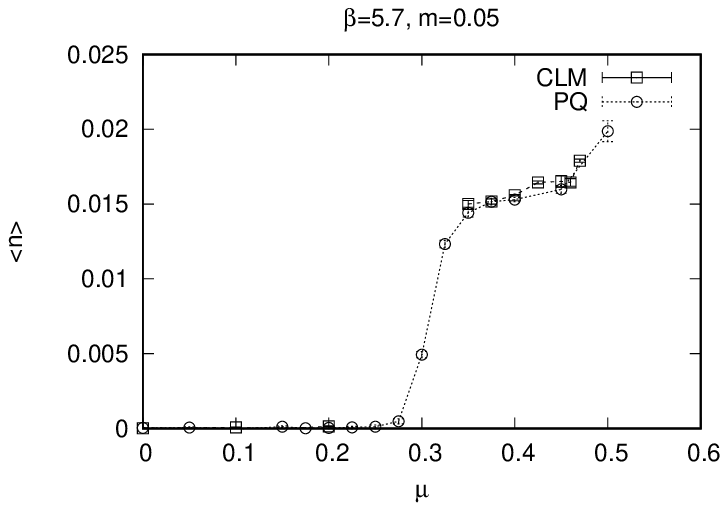}
\caption{The baryon number density is shown as a function
of the chemical potential $\mu$
for $m=0.01$ (Left) and $m=0.05$ (Right). 
The squares represent the CLM results which are reliable
judging from the probability distribution of the drift term.
The circles represent the RHMC results for the phase quenched model.
\label{fig:baryon}
}
\par\end{centering}
\end{figure}

Fig.\ \ref{fig:baryon} shows the baryon number density
\begin{align}
\langle n\rangle=\frac{1}{3N_{V}}\frac{\partial}{\partial\mu}\ln Z
\label{def-n}
\end{align}
obtained by the CLM for $m=0.01$ (Left) and $m=0.05$ (Right),
where we plot only the reliable
data judging from the probability distribution of the drift term.
We also plot the RHMC results for the phase quenched model for comparison.

As far as we can see from the reliable data,
the results obtained by the CLM do not depend much on the quark mass. 
Most strikingly, we observe
a clear plateau for $m=0.01$
in the region $0.325\lesssim\mu\lesssim0.425$,
which is visible for $m=0.05$ as well although it is less clear. 
We can also see some tendency of $\langle n\rangle$ starting to grow
gradually at larger $\mu$.
On the other hand, the results for 
the phase quenched model depend drastically on the quark mass
unlike the full model results.
In particular, for $m=0.01$ we observe a lower plateau
at $0.275 \lesssim \mu \lesssim 0.35$, 
which does not appear for $m=0.05$.

In order to understand these results, it is important to note that
the lattice spacing 
for $\beta=5.7$ and $m=0.01$ with
$N_{\mathrm{f}}=4$ staggered fermions is obtained 
as $a\sim0.045$ fm, while it is slightly larger for $m=0.05$.
Therefore the physical spatial extent of our lattice
is much smaller than the QCD scale
and the physical temperature is actually quite high.
The observed behavior which mimics that in the low temperature regime 
in a large volume may be understood as a result of the
temperature being lower than the scale of the spatial extent.

Another important thing to note is that 
the phase quenched model may be regarded as
a full model with $\mu$ being interpreted as the isospin chemical potential
due to $\det D(-\mu) = (\det D(\mu))^{*}$,
and in that case $\langle n\rangle$ represents 
$\langle n\rangle = \langle (n_{\rm u}-n_{\rm d})/3 \rangle$,
where $n_{\rm u}$ and $n_{\rm d}$ represent the ``up-quark'' density
and the ``down-quark'' number density, respectively.

The plateau in the phase quenched model
may therefore be identified as a state with 4 mesons, which corresponds
to $\langle n \rangle = 4 \times 2 / 3/8^3 = 0.0052$.
The reason why we observe the plateau instead of smooth increase
of the isospin density as is typically observed in previous 
work \cite{Kogut:2002zg}
is presumably due to the small physical volume of our lattice.
The plateau in the full model observed by the CLM, on the other hand,
may be identified 
as a state with 8 baryons, which corresponds
to $\langle n \rangle = 8 / 8^3 = 0.0156$.
The density of this state is much higher than what one would expect
for the nuclear matter, but this may be also due to the severe 
finite volume effects.

\begin{figure}
\begin{centering}
\includegraphics{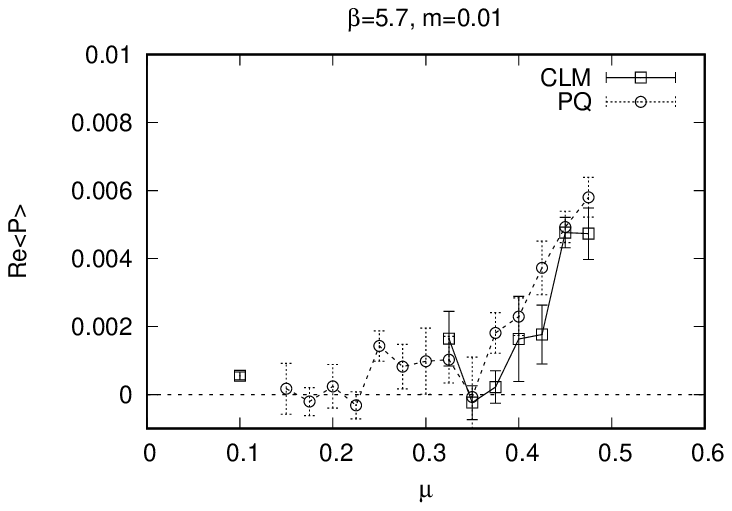}\includegraphics{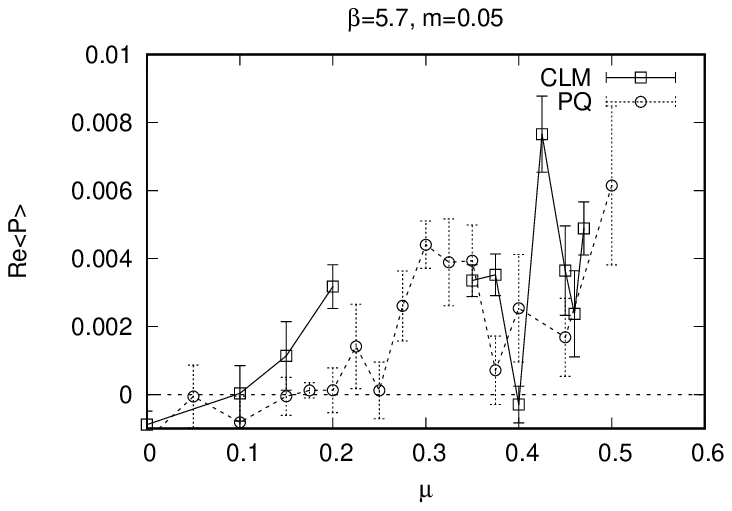}
\caption{The expectation value of the Polyakov loop 
is plotted against $\mu$ for $m=0.01$ (Left) and $m=0.05$ (Right).
The squares represent the CLM results which are reliable
judging from the probability distribution of the drift term.
The circles represent the RHMC results for the phase quenched model.
\label{fig:The-Polyakov-loop}}
\par\end{centering}
\end{figure}

In Fig.\ \ref{fig:The-Polyakov-loop}
we plot the expectation value of the Polyakov loop,
which is found to be small at $\mu\lesssim 0.45$
for both models with $m=0.01$ and $m=0.05$ despite the high temperature.
This is considered as
a consequence of the small spatial extent of our lattice.

\section{Summary and discussions\label{sec:Summary-and-discussions}}

In this paper we have made an attempt to
extend the success of the CLM in finite density QCD
to the lower temperature region with reasonably small quark mass
using four-flavor staggered fermions on a $8^3 \times 16$ lattice.
The physical size of our lattice is small due to the
choice $\beta=5.7$, which is motivated by the necessity to avoid 
the excursion problem \cite{Sinclair:2017zhn}.
With that in mind, our results can be summarized as follows.

First we have investigated the validity of the CLM based on the 
probability distribution of the drift term.
Interestingly, there is a region in which
the CLM works even at large $\mu$.
This is in contrast to the situation of the previous work 
on a $4^3 \times 8$ lattice \cite{Nagata:2018mkb},
where the use of the deformation technique \cite{Ito:2016efb}
was necessary to avoid the singular-drift problem in the large $\mu$ region.
In fact, when $\mu$ is chosen within the region of validity of the CLM,
the eigenvalue distribution of the Dirac operator exhibits
a gap along the real axis,
and therefore the singular-drift problem does not occur.

Second the baryon number density exhibits a plateau as a function 
of the chemical potential, which may be identified as the formation of
the nuclear matter. 
The gradual increase of the baryon number density starting at the end 
of the plateau suggests a continuous transition to the quark matter.
The comparison with the RHMC results for the phase quenched model
reveals a clear difference in the case of $m=0.01$.
These results encourage us to increase the lattice size further.
Simulations on a $16^4$ lattice are on-going.

\section*{Acknowledgements}

This research was supported by MEXT as
``Priority Issue on Post-K computer'' 
(Elucidation of the Fundamental Laws and Evolution of the Universe) 
and 
Joint Institute for Computational Fundamental Science (JICFuS).
Computations were carried out
using computational resources of the K computer 
provided by the RIKEN Advanced Institute for Computational Science 
through the HPCI System Research project (Project ID:hp180178).
J.~N.\ was supported in part by Grant-in-Aid 
for Scientific Research (No.\ 16H03988)
from Japan Society for the Promotion of Science. 
S.~S.\ was supported by the MEXT-Supported Program 
for the Strategic Research Foundation at Private Universities 
``Topological Science'' (Grant No.\ S1511006).

\bibliographystyle{h-physrev5}
\bibliography{ref}

\end{document}